\documentclass[runningheads]{llncs}

\usepackage{cite}
\usepackage{amsmath,amssymb,amsfonts}
\usepackage{algorithmic}
\usepackage{graphicx}
\usepackage{textcomp}
\usepackage{xcolor}
\def\BibTeX{{\rm B\kern-.05em{\sc i\kern-.025em b}\kern-.08em
    T\kern-.1667em\lower.7ex\hbox{E}\kern-.125emX}}

\usepackage{listings}

\usepackage{tikz}
\usepackage{float}
\usepackage{listings}
\usepackage{color}
\usepackage{textcomp}

\usepackage{multirow}

\usepackage{xcolor}
\usepackage{colortbl}

\usepackage{tabularx}
\usepackage{array}
\usepackage{varwidth} 

\usepackage{relsize}
\usepackage{xspace}
\PassOptionsToPackage{hyphens}{url}\usepackage{hyperref}

\hypersetup{pdftex,colorlinks=true,allcolors=blue}
\usepackage{hypcap}

\ifdefined\directlua
  \usepackage{fontspec}
\else
  \usepackage[T1]{fontenc}
  \usepackage[nomath]{lmodern}
\fi

\usepackage{ifthen}

\usepackage[normalem]{ulem}
\usepackage{keystroke}

\usepackage{subfig}
\usepackage{footnotebackref}

\usepackage{dblfloatfix}

\usepackage{fancyhdr}

\usepackage{caption}
\usepackage[frozencache,newfloat]{minted}

\providecommand{\e}[1]{\ensuremath{\times 10^{#1}}}

\newcommand\Cpp{C\nolinebreak[4]\hspace{-.05em}\raisebox{.4ex}{\relsize{-3}{\textbf{++}}}}

\definecolor{mygreen}{rgb}{0,0.6,0}
\definecolor{myred}{rgb}{0.6,0,0}

\definecolor{mygray}{rgb}{0.5,0.5,0.5}
\definecolor{mymauve}{rgb}{0.58,0,0.82}
\usetikzlibrary{shapes,positioning}

\newcolumntype{M}{>{\begin{varwidth}{18mm}}l<{\end{varwidth}}} 

\fancypagestyle{pprintTitle}{%
\lhead{}
\chead{\vspace{-20mm}Accepted in \href{https://2021.euro-par.org/}{27$^{th}$ Int. European Conf. on Parallel and Distributed Computing (EuroPar 2021)}. 30 August-3 September 2021, Lisbon, Portugal\vspace{10mm}}%
\lfoot{}\cfoot{}\rfoot{}

}

\newcommand{\withSpacedSections}{false}
\newcommand{\maybeSpacedSection}{\ifthenelse{\equal{\withSpacedSections}{true}}{\clearpage}{}}

\newcommand{\maybeDelOption}{highlight}

\newcommand{\modeArXiV}{yes}

\newcommand{\maybeDel}[2]{%
 \ifthenelse{\equal{\maybeDelOption}{yes}}{
     #2 
   }{
   \ifthenelse{\equal{\maybeDelOption}{no}}{%
     #1
   }
   {%
    {\color{myred}\sout{#1}}
    {\color{mygreen}#2}
   }%
 }%
}

\newcommand{\ifArXiV}[1]{%
 \ifthenelse{\equal{\modeArXiV}{yes}}{
 #1
 }{}
}

\begin{document}

\title{Exploiting co-execution with oneAPI: heterogeneity from a modern perspective}
\titlerunning{Exploiting co-execution with oneAPI}%


\author{Raúl Nozal \and
Jose Luis Bosque}
%
\authorrunning{R. Nozal \and JL. Bosque}
%
\institute{Department of Computer Science and Electronics, Universidad de Cantabria, Spain\\
\email{\{raul.nozal,joseluis.bosque\}@unican.es}}


\maketitle\ifArXiV{\thispagestyle{pprintTitle}}



\begin{abstract}
Programming efficiently heterogeneous systems is a major challenge, due to the complexity of their architectures. Intel oneAPI, a new and powerful standards-based unified programming model,  built on top of SYCL, addresses these issues. In this paper, oneAPI is provided with co-execution strategies to run the same kernel between different devices, enabling the exploitation of static and dynamic policies. On top of that, static and dynamic load-balancing algorithms are integrated and analyzed.
This work evaluates the performance and energy efficiency for a well-known set of regular and irregular HPC benchmarks, using an integrated GPU and CPU. Experimental results show that co-execution is worthwhile when using dynamic algorithms, improving efficiency even more when using unified shared memory.

\keywords{Heterogeneous computing \and parallel computing \and co-execution \and load balancing \and SYCL \and oneAPI \and DPC++ \and scheduling \and HPC}
\end{abstract}


\section{Introduction}
\label{sec:Introduction}



The future of computing cannot be understood without heterogeneous computing \cite{Zahran:17}, due to its excellent cost/performance ratio and energy efficiency. This facilitates the acceleration of a wide range of massively data-parallel applications, such as deep learning \cite{Lin:2020}, video processing \cite{Costero:2020,ZernikeJPDC:12} 
or financial applications \cite{Castillo:2015}.  
However, hardware heterogeneity complicates the development of efficient and portable software, especially when specialized components require their own programming models. In this context, some of the hot topics being researched are: supporting single source programming, improving the usability and efficiency of memory space, distributing computation and data among different devices, and load balancing \cite{Zhang:2017, Shen:2016,Beri:2017,Nozal:20,Shin:2020,maat,Perez:17,nozal2019load}.

Programming models have become more abstract and expressive. 
OpenCL emerged as an open standard programming model for writing portable programs across heterogeneous platforms \cite{opencl:Gaster:2013}. However, it has a very low level of abstraction and leaves to programmers the partitioning and transferring of data and results among the CPU and devices. 
On the other end, proposals based on compiler directives have been developed, such as OpenACC \cite{Farber:2016}, and later extensions of OpenMP \cite{Vitali:2019}, leaving all this work to the compiler, but limiting both the expressiveness and performance.
%
%
Moreover, market trends and industrial applications indicate a strong predominance of languages such as C++, favoring higher level alternatives. For instance, SYCL is a cross-platform abstraction layer that builds on OpenCL, enabling the host and kernel code to be contained in the same source file with the simplicity of a cross-platform asynchronous task graph \cite{SYCL:2020}.

In this context, Intel has developed oneAPI, a unified programming model to facilitate the development among various hardware architectures \cite{OneAPI:2020}. It provides a runtime, a set of domain-focused libraries and a simplified language to express parallelism in heterogeneous platforms. It is based on industry standards and open specifications, offering consistent tooling support and interoperability with existing HPC programming models. The oneAPI's cross-architecture language Data Parallel C++ (DPC++) \cite{Ashbaugh:20}, based on SYCL standard for heterogeneous programming in C++, provides a single, unified open development model for productive heterogeneous programming and cross-vendor support. It allows code reuse across hardware targets, while permitting custom tuning for a specific accelerator. 
Some of the features provided comprise optimized communication patterns, automatic dependency tracking, runtime scheduling and shared memory optimizations, between others.

This article addresses a new challenge in improving the usability and exploitation of heterogeneous systems, providing oneAPI with the capacity for \emph{co-execution}. This is defined as the collaboration of all the devices in the system (including the CPU) to execute a single massively data-parallel kernel \cite{Zhang:2017, Shen:2016, Nozal:20, nozal2019time}. However, it is a hard task for the programmer and needs to be done effortless in order to be widely used. In this way, the expression and abstraction capabilities of oneAPI, such as portability and single-source style, will be exploited to obtain codes that will be easier to implement and maintain. To efficiently exploit the computing capacity of all devices, a series of workload balancing algorithms are implemented, both static and dynamic, obtaining good results with both regular and irregular applications.
%
Experimental results show that co-execution is worthwhile from the point of view of performance and energy efficiency as long as dynamic schedulers are used, and even more if unified memory is applied.


Although oneAPI release is very recent, it has quickly attracted the attention of industry and the scientific community working with heterogeneous systems. A SYCL-based version of the well-known Rodinia benchmark suite has been developed in \cite{jin2020rodinia}, using Intel oneAPI toolkit. 
Christgau and Steinke \cite {christgau:20} use both the compatibility tool {\em dpct} of oneAPI, as well as SYCL extensions for the CUDA base code of the easyWave simulator. 
A study of the performance portability between different Intel integrated GPUs using oneAPI is presented in \cite{zheming:19}, where a computationally intensive routine is derived from the Hardware Accelerated Cosmology Code (HACC) framework. 
A debugger based on GDB for SYCL programs that offload kernels to CPU, GPU, or FPGA emulator devices, has been developed as part of the oneAPI distribution \cite{aktemur:20}. 

As far as we know, the only work that addresses co-execution with oneAPI is \cite{asenjoJOS:20}. The authors extend the Intel TBB \emph{parallel\_for} function to allow simultaneous execution of the same kernel on CPU and GPU. They implement three schedulers on top of oneAPI, static, dynamic and adaptive LogFit. The main differences with our work are that we provide a pure oneAPI architecture (without TBB) and present a rich variety of kernels, both regular and irregular, which reveal differences in the behavior of schedulers. 

The rest of the paper is organized as follows. 
Section \ref{sec:Motivation} describe the issues that motivate this work while in Section \ref{sec:Coexecution}, the co-execution architecture and its design decisions are exposed. The methodology used for the validation is explained in Section \ref{sec:Methodology}, while the experimental results are shown in Section \ref{sec:Validation}. Finally, Section \ref{sec:Conclusions} highlights the most important conclusions and future work.

\section{Motivation}
\label{sec:Motivation}


Intel oneAPI uses the host-device programming model, where the host offloads compute-heavy functions, called kernels, to a set of hardware accelerators, such as GPUs and FPGAs. Its runtime is able to manage complex applications composed of a set of kernels, even if they have dependencies between them, through a Directed Acyclic Graph (DAG). The assignment of a kernel to a particular device can be done by the programmer, so it is determined at compile time, or let oneAPI choose the device at runtime. In either case, a kernel can only be scheduled to a single device when the dependencies are satisfied.

In this context, the only possibility of co-execution is for the programmer to split the work into several kernels, as many as there are devices in the system. Also, data partition and workload distribution must be done manually. Furthermore, the compiler must detect that these kernels are independent and schedule them simultaneously.
This complicates the co-execution and, therefore, the exploitation of the whole system to solve a single kernel.

Even if the programmer is willing to face this extra effort, an additional problem arises with workload balancing. Since the division of the workload is done at compile time, it is necessarily static. That is, the portion of work assigned to each device is pre-fixed at the beginning of execution. This partitioning works well for regular applications, where the execution time of a data set depends only on its size \cite{Zhang:2017}. The programmer needs to estimate off-line how much workload to allocate to each device so that both finish at the same time, thus obtaining a balanced execution, as seen in the left part of Figure \ref{fig:static-irregular}, for the Gaussian kernel. In this case the kernel execution time is 5 seconds on the CPU and 2 second on the GPU, which means that the GPU has 2.5x the performance of the CPU. Therefore, assigning the work to devices proportionally to their computing capabilities, a balanced distribution is obtained and the execution time is reduced to approximately 1.5 seconds. 

\begin{figure}[h]
\centering
   \includegraphics[width=1.0\columnwidth, keepaspectratio]{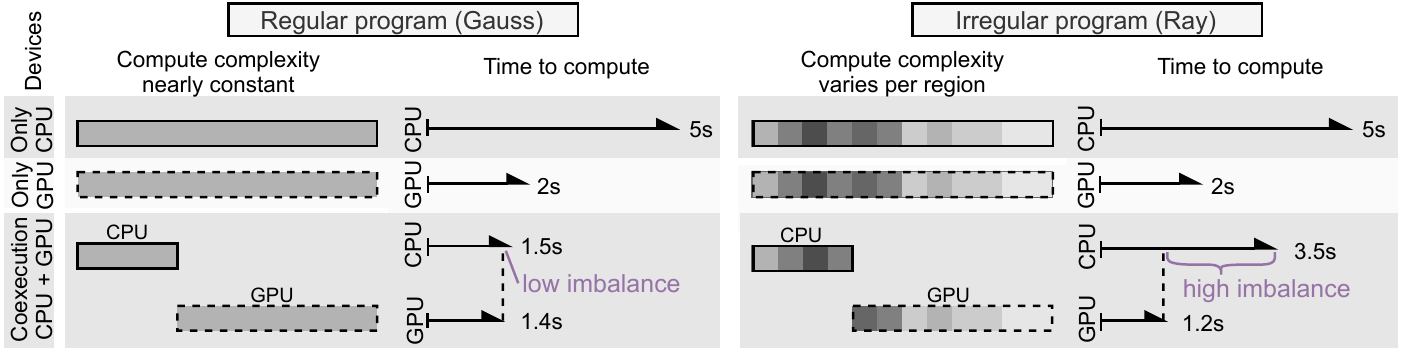}
   \vspace{-4mm}
   \caption{Static co-execution for regular and irregular programs.}
   \label{fig:static-irregular}
\end{figure}

However, it is well known that static scheduling cannot adapt to the irregular behavior of many applications, leading to significant load imbalances \cite{Nozal:20}. In these applications, the processing time of a data set depends not only on its size, but also on the nature of the data. Thus, different portions of data of the same size can generate different response times. This is shown in the right part of Figure \ref{fig:static-irregular}, which presents the execution of a Raytracer application on two devices. The darker shades of grey refer to more computationally intensive data areas. Performing the same static balancing as in the regular case, it has coincided that the most computationally heavy regions have fallen on the CPU (slower device). This resulted in a significant imbalance, with the CPU taking 3.5 seconds while the GPU took only 1.2 seconds. This situation can only be addressed with dynamic balancing algorithms that allocate portions of work to the devices on demand.

This paper addresses both of these problems. On the one hand, it is proposed to provide oneAPI with mechanisms that allow the implementation of co-execution without additional effort for the programmer. On the other hand, it provides the oneAPI scheduler with a set of dynamic load balancing algorithms to squeeze the maximum performance out of the heterogeneous system, even with irregular applications.

\maybeSpacedSection
\section{Co-execution based on OneAPI}
\label{sec:Coexecution}

The approach to achieve co-execution focuses on using the DPC++ compiler and runtime, hereafter referred to as oneAPI for simplicity. The proposed {\em Coexecutor Runtime} is built on top of oneAPI as a runtime library to allow the parallel exploitation of multiple hardware accelerators that facilitate the implementation of workload balancing algorithms.

This approach has several architectural and adaptive advantages.  Firstly, the design and implementation are based on open standards, both C++ and SYCL, following easily recognizable architectural patterns.
Secondly, since it is drawing on previous standards such as OpenCL, it facilitates the adaptation for a whole repertoire of libraries and software generated over a decade, helping to benefit from co-execution. Thirdly, it serves as a skeleton upon which to apply different strategies and workload balancing algorithms for using oneAPI and SYCL. Finally, as it is designed from a sufficiently standardized and abstract approach, it allows the adaptation and extension to execution engines and proposals created by other manufacturers, both compilers and accelerator drivers.

To provide oneAPI with co-execution depends mainly on the correct detection of a potential concurrent execution path by the compiler and the runtime. This materializes a parallel execution of several tasks of the DAG, thanks to the existence of totally independent hardware resources.
To achieve good results, it is necessary to use dynamic strategies, currently not available in oneAPI. In addition, it is important to implement workload balancing algorithms to obtain the best possible performance. Both aspects are explained below.


\subsection{Dynamic co-execution}

The strategy proposed for dynamic co-execution is to promote multithreaded management architectures based on the runtime of oneAPI.
The {\em Coexecutor Runtime} enhances the isolation between execution devices, since one of the key points is to make it easier for the compiler to detect disjoint memory structures as well as the independence between queues and tasks.
In addition, since oneAPI offers a sufficiently sophisticated and complete memory model, the management architecture must be adapted to favor both buffer management and the possibility of exploiting unified memory (USM).

To define the proposal, three perspectives are considered, the execution model, from the memory point of view and the last one, the relationship of the {\em Coexecutor Runtime} with the runtime of oneAPI, as it is explained in Section \ref{sec:CoexecutionLBA}.

\begin{figure*}[!b]
    \centering
    \subfloat[Execution model as part of a blocking section of an application.\label{fig:execution-memory-models-exec}]{\includegraphics[width=0.44\textwidth, keepaspectratio]{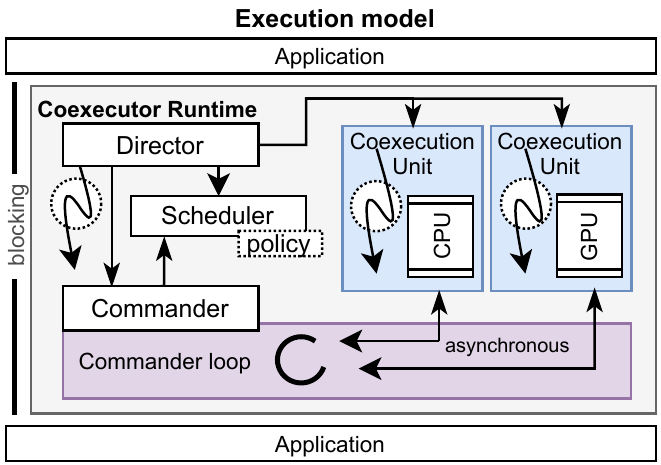}}%
    \hfill
    \subfloat[Memory model example for USM and SYCL buffers when using CPU and GPU.\label{fig:execution-memory-models-mem}]{\includegraphics[width=0.54\textwidth, keepaspectratio]{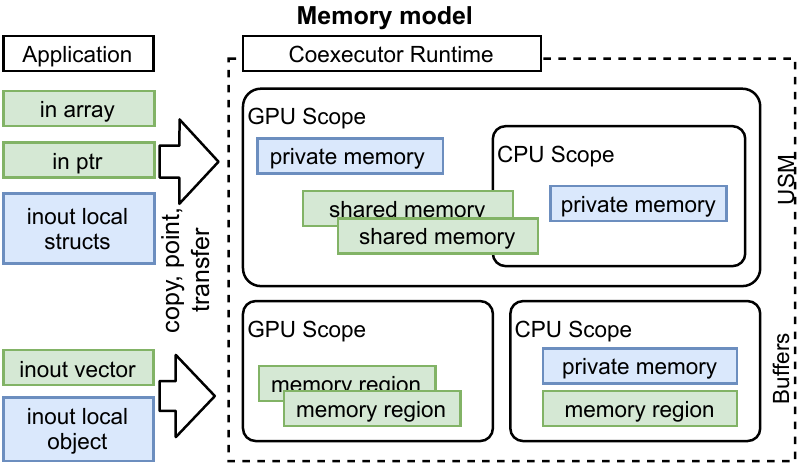}}%
    \caption{Coexecutor Runtime considering CPU-GPU dynamic co-execution.}
\end{figure*}


The {\em execution model} is shown in Figure \ref{fig:execution-memory-models-exec}, representing the interaction of the runtime as part of the execution process of an application.
Execution is blocked from an application point of view, although internally it works asynchronously.

The {\em Director} configures the {\em Coexecution Units} and manages both the {\em Commander} and its communication with the rest of the entities. The {\em Scheduler} is instantiated and plugged in with a policy established by the programmer, using one of the schedulers explained in Section \ref{sec:CoexecutionLBA}. The {\em Commander} is responsible of packaging the work, emitting tasks and receiving events, as part of the computation workflow with the {\em Coexecution Units}. This process is termed as {\em Commander loop}, and it follows the scheduling strategy defined by the {\em Scheduler}.

Regarding the {\em Coexecutor Runtime} internal workflow, the {\em Director} instantiates and configures oneAPI primitives and structures necessary both for the operation with oneAPI runtime and used by the {\em Scheduler} itself, among which are work and queue entities, execution contexts and mapping of memory structures between the application and the runtime.
In parallel, the management threads of the {\em Coexecution Units} initialize the communication mechanisms within the runtime, as well as the request of devices and their configuration with oneAPI.
The communication is bidirectional between {\em Commander} and each {\em Coexecution Unit}, since it is co-executed with an independent scheduler that handles the decisions. As soon as there is a {\em Coexecution Unit} ready to receive work and the management thread has finished the initial phase, it establishes communication with the {\em Commander loop}. As the rest of the devices are completing their initialization, they incorporate into the loop, where the scheduling phase starts.


\begin{figure*}[!b]
    \centering
    \includegraphics[width=0.99\columnwidth, keepaspectratio]{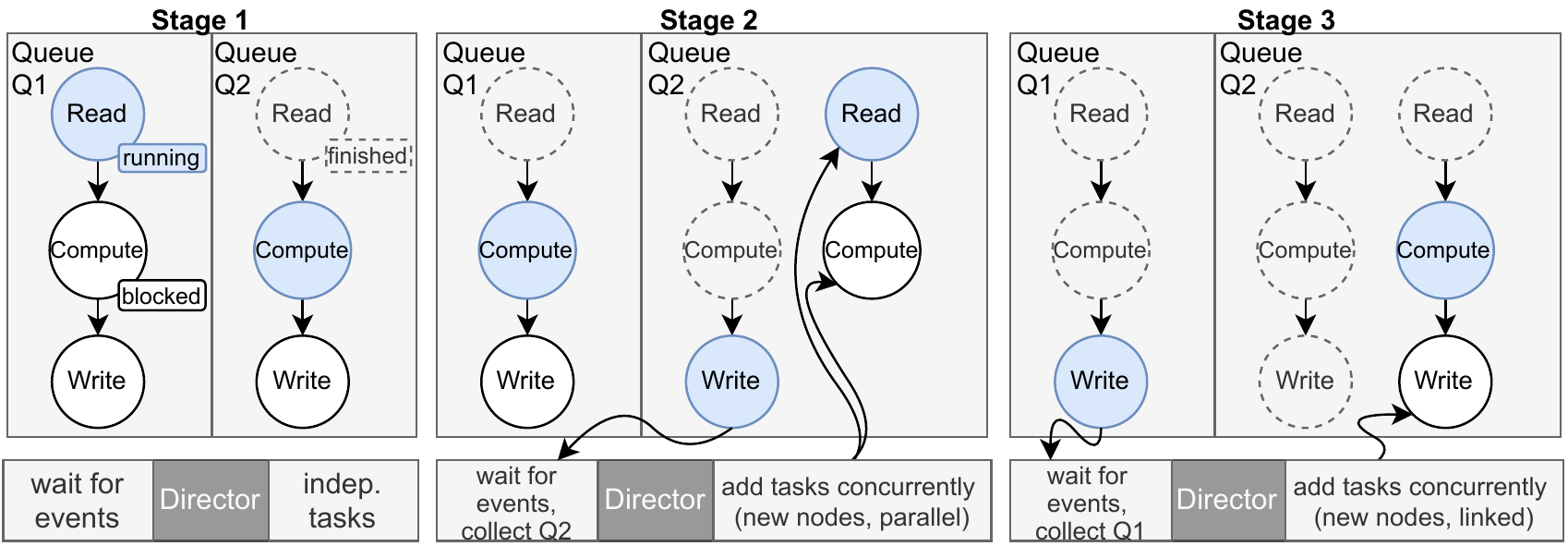}
    \vspace{-2mm}
    \caption{Example of interaction with the DAG from oneAPI's perspective while running a dynamic approach with two queues.}
    \label{fig:dynamic-coexecution}
\end{figure*}

The {\em memory model} is presented in Figure \ref{fig:execution-memory-models-mem}. It shows the separation between structures and memory containers, taking into account the two types of strategies used: USM or buffers of SYCL, although the {\em Coexecutor Runtime} supports the combination of both during the co-execution.
On the left of the figure are shown the structures, \Cpp{} containers and memory pointers used by the application, while the right outlines the view of the runtime.
The {\em Director} and its {\em Coexecution Units} handle the allocations and configuration of the memory space with oneAPI, and the programmer only has to request the use. The runtime will distribute them in the oneAPI memory model, either by transferring pointers, copying memory regions or sharing unified memory blocks.



Two ways of operating with oneAPI memory environments are distinguished.
If USM is used, the {\em Coexecutor Runtime} provides two scopes: a larger (upper) one for a device (GPU) and a smaller (lower) for another (CPU). 
This way, the memory spaces initialized by the GPU are reused in the CPU using oneAPI primitives. On the other hand, if SYCL buffers are used, the scope of each device will manage independent buffers with memory regions that will be part of a higher container or structure. Therefore, favoring the recognition of disjointed memories by the compiler. Private memory allocations can be made in both memory models, in the form of buffers and variables, where each field is controlled independently by each {\em Coexecution Unit} and its oneAPI's scopes. 

Finally, both ways of operating can be combined, since could be parts that use the USM model and others that rely on buffers and variables. {\em Coexecutor Runtime} will reuse the scope of each device to map any C++ containers and memory sections, each of which will be governed by a memory model.

The interaction between the {\em Coexecutor Runtime} and oneAPI is shown in Figure \ref{fig:dynamic-coexecution}. Three stages are presented during the execution of the runtime, with two different queues Q1 and Q2. It starts from a situation where the runtime has established two independent parallel execution queues, due to the existence of two separate underlying architectures. The nodes of each queue are managed by the runtime and its DAG, and they can be in execution (blue), blocked waiting for resources (white) or finished (gray with a dashed line). The {\em Director} waits for events related to the DAG or performs independent tasks, such as resource management, receiving and sending notifications, status control or work reparation, some of which are essential within the {\em Scheduler}.

By switching to the stage 2, it can be distinguished how the Q2 is able to process nodes more efficiently, so the {\em Director} collects results of the write operation and enqueues new nodes of the DAG to the same queue, overlapping computation and communication. Collection operations are dependent on the memory model, the type of operations (explicit or implicit) and the amount of bytes used, thus they could be fast (unified memory) or slow (mixed models or transfer large blocks). Finally, in the third stage, the end of the Q1 is represented with the output data collection while in the Q2 a next writing task is added. This is linked to the branch created in stage 2, as soon as its computation task has started, distributing the DAG management among different time periods.





\subsection{Load balancing algorithms}
\label{sec:CoexecutionLBA}

\begin{figure*}[!b]
    \centering
    \includegraphics[width=0.95\columnwidth, keepaspectratio]{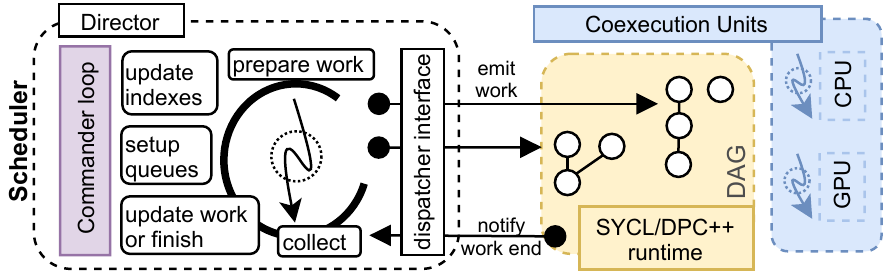}
    \vspace{-2mm}
    \caption{Commander's loop where the scheduling strategy is performed to coordinate the behaviors of the Coexecution Units.}
    \label{fig:dyn-scheduler}
\end{figure*}

To enable dynamic policies to squeeze all the computing capacity out of the heterogeneous system, the {\em Scheduler} component is introduced, as it is shown in Figure \ref{fig:execution-memory-models-exec}. It configures the behavior of the load balancer, the distribution and division of the work packages, as well as the way to communicate with the different execution devices.

Figure \ref{fig:dyn-scheduler} depicts the relationship of the {\em Coexecutor Runtime} with the runtime of oneAPI, all of it involved as part of the {\em Commander loop}. The {\em Coexecutor Runtime} internal communication is performed between the management threads, either those associated with the devices (right) or the global manager, usually associated with the {\em Director} (left). This view simplifies the runtime of oneAPI and its internal DAG management, being considered as a single entity, part of the {\em Coexecution Units} (right). The {\em Director} performs a set of periodic actions, as a loop managing events and operations, among which are: preparing the next job to be issued; collecting completed jobs; updating pending jobs; 
preparing and reusing the queue and command groups as well as other oneAPI primitives; and updating the indexes, ranges and offsets of memory entities.

Every time a work package is prepared, the runtime adds a task in the DAG. Similarly, with the completion of a job, {\em Commander} receives the notification to collect and merge the output data, if needed. This operation can be lightweight in case of using USM or using implicit operations, delegating more responsibility to oneAPI. The emission and reception of work is requested through a dispatch interface, as a way of unifying requests. Finally, when there are no more pending jobs, the {\em Commander} will notify the {\em Director} to close and destroy the primitives and management objects to return control to the application.

As a result of the proposed architecture based on the designed dynamic co-execution model, three algorithms are implemented in the {\em Scheduler} \cite{nozal2019time,maat,Nozal:20}. The {\em Static} algorithm divides the kernel in as many packages as devices are in the system, minimizing the number of host-devices interactions. The size of each package is proportional to the relative computing speed of each device. Its drawbacks are that it is difficult to find a suitable division and that cannot adapt its behavior dynamically to irregular applications.
It has a minimum management inside the {\em Scheduler} component, because it only runs as many iterations in the loop of events as devices are co-executing.

Regarding the strictly dynamic strategies, the {\em Dynamic} algorithm divides the data in a number of packages of similar size, which are assigned to the devices on demand, as soon as they are idle. This allows it to adapt to the irregular behavior, but increments the overhead communication between host and devices. On the other side, {\em HGuided} starts with large packages and decreases their size as the execution progresses. The size of the initial packages is proportional to the computing capacity of the devices. Therefore, it reduces the number of synchronization points while retaining most of its adaptiveness. Considering these dynamic policies inside the {\em Scheduler}, it is not possible to know in advance the quantity of iterations, because it will depend on each execution parameters, as well as the number and type of devices. These operations increase the management overhead due to the operations related to the update of indexes and ranges, as well as the division of the problem into independent regions. Finally, concerning the differences in the operations carried out by {\em Commander}, {\em Dynamic} will simplify the number of instructions involved in the calculation of work packages compared to {\em HGuided}. This is explained since the latter performs a more sophisticated algorithm that takes into account certain conditions, including the computing power of each device. However, the calculation overheads of the latter are compensated by the efficiency of its workload distribution policy.

\subsection{API design}
\label{sec:api}

\begin{listing}[!t]

  \begin{minted}[mathescape,
    linenos,
    numbersep=5pt,
    gobble=2,
    frame=lines,
    breaklines,
    framesep=2mm]{c++}
   coexecutor_runtime<hg> runtime;
   runtime.config(CounitSet::CpuGpu, coexecutor_runtime::dist(0.35));
   runtime.launch(data.size(), [&](coexecutor_unit *counit, package pkg) {
     sycl::buffer<int, 1> buf_input(data.data() + pkg.offset,
                                    sycl::range<1>(pkg.size));
     counit->dispatch([&](sycl::handler &h) {
       auto R = sycl::range<1>(pkg.size);
       auto input = buf_input.get_access<sycl::access::mode::read_write>(h);
       h.parallel_for(R, [=](sycl::item<1> it) {
         auto tid = it.get_linear_id();
         input[tid] = input[tid] * datav;
       });
     });
   });
  \end{minted}
  \vspace{-5mm}
  \captionof{listing}{Coexecutor Runtime computing SAXPY with a dynamic algorithm using simultaneously CPU and GPU.}
  \label{lst:saxpy}
\end{listing}

Coexecutor Runtime has been designed to offer an API that is  flexible as well as closely linked to the SYCL standard, favoring reuse of existing code and slightly higher usability. Listing \ref{lst:saxpy} shows a simple example of use when computing the SAXPY problem simultaneously exploiting both CPU and GPU. The code fragment where the runtime is used is shown, so the initialization of the problem and its data, as well as the subsequent usage, are omitted.

Line 1 instantiates the {\em coexecutor\_runtime} prepared to compute a program using the HGuided balancing algorithm. In the next line, it is configured to use both the CPU and GPU, giving a hint of the computational power of 35\% for the CPU in proportion to the GPU. This value will leverage the algorithm to further exploit co-execution efficiency. Next, the co-execution scope associated with the problem is provided (lines 3 to 14), where a lambda function captures by reference the values used. This scope is executed by each of the {\em Coexecution Units}, and therefore, they must establish independent memory reservations (or shared, if shared virtual memory is exploited), using the values provided by the runtime itself through the {\em package} class. Line 6 opens an execution scope, associated to the kernel computation for each device. In lines 7 and 8 a read and write access is requested for the previous memory region (buffer {\em accessors}), indicating the execution space based on the given package size. Finally, lines 9 to 11 show the data-parallel execution, traversing the indicated execution space ({\em R}) and using the {\em accessors} and variables needed ({\em datav, input}).

In the line immediately following what is shown in the example, the problem will have been computed simultaneously using both devices. In addition, the data resulting from the computation will be in the expected data structures and containers (vector {\em input} of C++).

\maybeSpacedSection
\section{Methodology}
\label{sec:Methodology}


The experiments to validate {\em Coexecutor Runtime\footnote{\label{footnote-github}
https://github.com/oneAPI-scheduling/CoexecutorRuntime}} have been carried out in a computer with an Intel Core i5-7500 Kaby Lake architecture processor, with 4 cores and an integrated GPU Intel HD Graphics 630. The GPU is a Gen 9.5 GT2 IGP, with 24 execution units. An LLC cache of 6 MB is shared between CPU and GPU.

To accomplish the validation, 6 benchmarks have been selected, which represent both regular and irregular behavior, as described in Section \ref{sec:Motivation}.
{\em Gaussian, MatMul} and {\em Taylor} correspond to regular kernels, while {\em Mandelbrot, Rap} and {\em Ray Tracing} are irregular ones. Taylor, Rap and Ray are open source implementations, while the rest belong to the AMD APP SDK, being all ported to oneAPI. 
Table \ref{tbl:benchs-properties} presents the most relevant parameters of the benchmarks, providing enough variety to validate the behavior of the runtime.

To guarantee integrity of the results, the values reported are the arithmetic mean of 50 executions, discarding a previous first one to avoid warm-up penalties. The standard deviation is not shown because it is negligible in all cases.


\begin{table}[!t]
    \begin{center}
        \caption{Benchmarks and their variety of properties.}
        \label{tbl:benchs-properties}
        \scriptsize
        \renewcommand{\arraystretch}{1.5}
        \begin{tabular}{%
@{\hskip 1mm}p{2.7cm}@{\hskip 2mm}%
@{\hskip 2mm}>{\columncolor[gray]{0.95}}c@{\hskip 2mm}%
@{\hskip 2mm}c@{\hskip 2mm}%
@{\hskip 2mm}>{\columncolor[gray]{0.95}}c@{\hskip 2mm}%
@{\hskip 2mm}c@{\hskip 2mm}%
@{\hskip 2mm}>{\columncolor[gray]{0.95}}c@{\hskip 2mm}%
@{\hskip 2mm}c@{\hskip 2mm}%
}
            \textbf{Property}
            & \rotatebox[origin=s]{0}{\centering\textbf{Gauss}}
            & \rotatebox[origin=c]{0}{\centering\textbf{Matmul}}
            & \rotatebox[origin=c]{0}{\parbox[c]{13mm}{\centering\textbf{Taylor}}}
            & \rotatebox[origin=c]{0}{\parbox[s]{13mm}{\centering\textbf{Ray}}}
            & \rotatebox[origin=c]{0}{\parbox[s]{13mm}{\centering\textbf{Rap}}}
            & \rotatebox[origin=c]{0}{\centering\textbf{Mandel}\vspace{3mm}}
            \\ \hline

            Local Work Size    & 128    & 1,64    & 64     & 128    & 128     & 256     \\[1pt] \hline
            Read:Write buffers & 2:1    & 2:1     & 3:2    & 1:1    & 2:1     & 0:1     \\[1pt] \hline
            Use local memory   & no     & yes     & yes     & yes     & no      & no      \\[1pt] \hline
            Work-items (N\e{5})        & 262 & 237   & 10     & 94     & 5   &  703    \\[1pt] \hline
            Mem. usage (MiB) & 195    & 264     & 46     & 35     & 6       & 1072    \\[1pt] \hline




        \end{tabular}
    \end{center}
\end{table}

The validation of the proposal is done by analyzing the co-execution when using four scheduling configurations in the heterogeneous system.
%
As summarized in Section \ref{sec:CoexecutionLBA}, Static, Dynamic and HGuided algorithms are evaluated, labelled as St, Dyn and Hg, respectively. In addition, the dynamic scheduler is configured to run with 5 and 200 packages. 
Finally, two different memory models have also been tested: unified shared memory (USM) and SYCL's buffers (Buffers). 


To evaluate the performance of the runtime and its load balancing algorithms, the total response time is measured, including kernel computing and data transfer. Then, two metrics are calculated: imbalance and speedup.
The former measures the effectiveness of load balancing, calculated as $\frac{ {T_{GPU}} }{ {T_{CPU}} }$, where $T_{GPU}$  and $T_{CPU}$ are the execution time of each device.
The speedup is computed as $S=\frac{T_{GPU}}{T_{co-exec}}$, because the GPU is the  fastest device for all the benchmarks.

%

Finally, energies are measured using RAPL counters, giving the total consumption in $Joules$. The metric used to evaluate the energy efficiency is the Energy-Delay Product (EDP).



\maybeSpacedSection
\section{Validation}
\label{sec:Validation}



\subsection{Performance}






The imbalances and speedups achieved with CPU-GPU co-execution are shown in Figure \ref{fig:imbalance-speedup}.
The abscissa axes show the benchmarks, each one with four scheduling policies and two memory models, as defined in Section \ref{sec:Methodology}. Moreover, the geometric mean for each scheduling policy is shown on the right side. Regarding balancing efficiency, the optimal is $1.0$, where both devices finish simultaneously without idle times. Any deviation from that value means more time to complete for one device compared with the other. Generally, the imbalance is below 1.0 due to the overheads introduced by the CPU when computing, as a device, and managing the runtime resources, as the host. It rarely completes its computation workload before the GPU finishes, since the latter requires more resource management by the host, increasing the CPU load.

The main conclusion that is important to highlight is that co-execution is always profitable from a performance point of view, as long as it is done with dynamic schedulers, and even more if using unified memory ({\em USM}), as the geometric mean summarizes for these benchmarks and scheduling configurations.

\begin{figure*}[!t]
    \centering
    \includegraphics[width=0.99\textwidth, keepaspectratio]{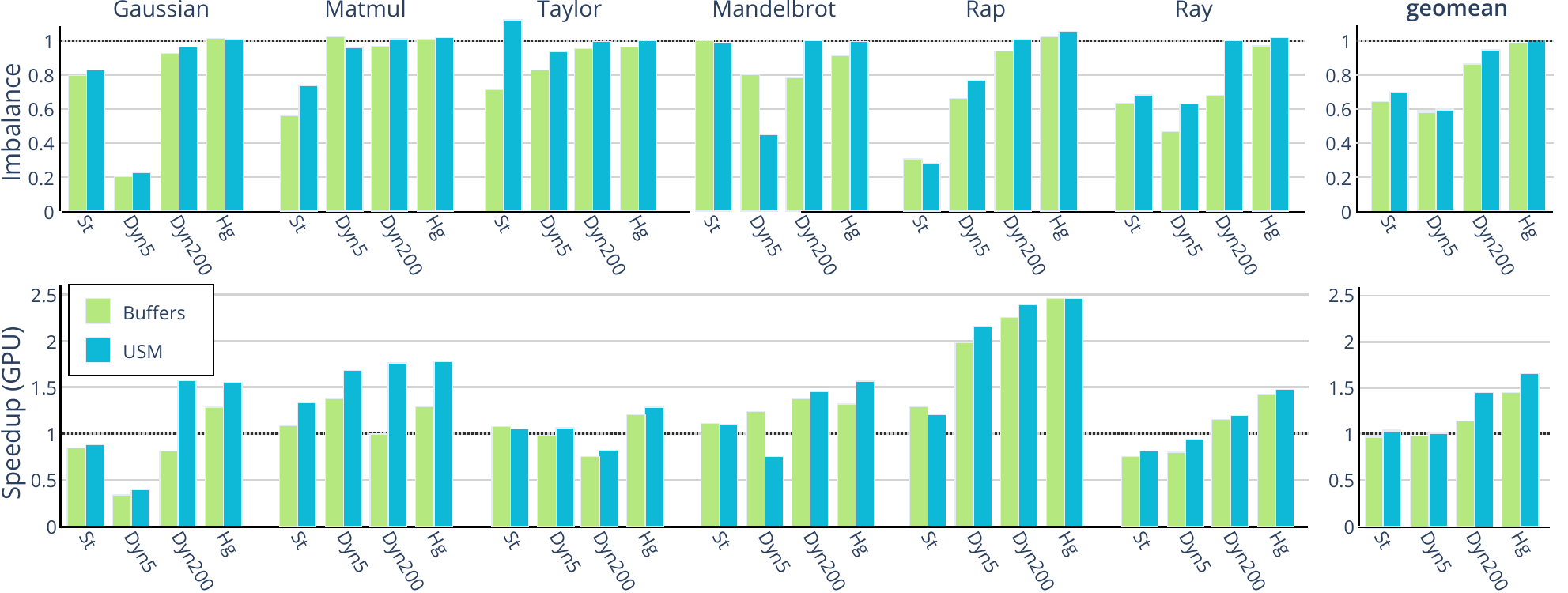}
    \vspace{-3mm}
    \caption{Balancing efficiency (top) and speedups (bottom) for a set of benchmarks when doing CPU-GPU co-execution.}
    \label{fig:imbalance-speedup}
\end{figure*}

Analyzing the different load balancing algorithms, it can be seen that the Static offers the worst performance, even in regular applications where it should excel.
This is because the initial communication overhead caused by sending a large work package, leads to a significant delay at the beginning of the execution, strongly penalizing the final performance.


%
%
Regarding dynamic algorithms, they provide good results in general, especially when the {\em USM} memory model is used. However, they have the drawback that the number of packages for each benchmark has to be carefully selected. A very small number of packages can lead higher imbalances causing a performance penalty, as can be seen in Gaussian, Mandelbrot or Ray, in the case of Dyn5. At the other extreme, a very large number of packages increases the communication overhead, impacting negatively on performance, as in Gaussian with {\em Buffers}. In between, there is a tendency that the greater the number of packages, the better the balancing.
This is an expected behavior because the packages are smaller and their computation is faster, giving less chance of imbalance in the completion of both devices. This is an interesting behavior since the {\em Coexecutor Runtime} is delivering high performance when using dynamic strategies due to the low overhead of the {\em Commander loop} when managing packages and events. 

The HGuided algorithm offers the best scheduling policy, thanks to its balancing efficiency near to 1. It yields the best performance in all the analyzed benchmarks, with speedups values ranging from 2.46 in Rap to 1.48 in Ray. Moreover, it does not require any a priori parameters, which simplifies its use for the programmer.



Considering the memory models, there is a general improvement in balancing and performance when using {\em USM} compared with {\em Buffers}. It can be observed than {\em USM} performs much better than {\em Buffers} on regular kernels and with dynamic strategies, but this difference practically disappears on irregular kernels.

Finally, it is important to highlight the relationship between the imbalance and the speedups obtained. Although generally less imbalance indicates better performance, this does not have to be the case if the imbalance is not very high and more amount of work has been computed by the faster device. This is the case of Ray when using {\em Buffers}, since more work is computed by the GPU.


\subsection{Energy}

\begin{figure*}[!t]
    \centering
    \includegraphics[width=0.99\textwidth, keepaspectratio]{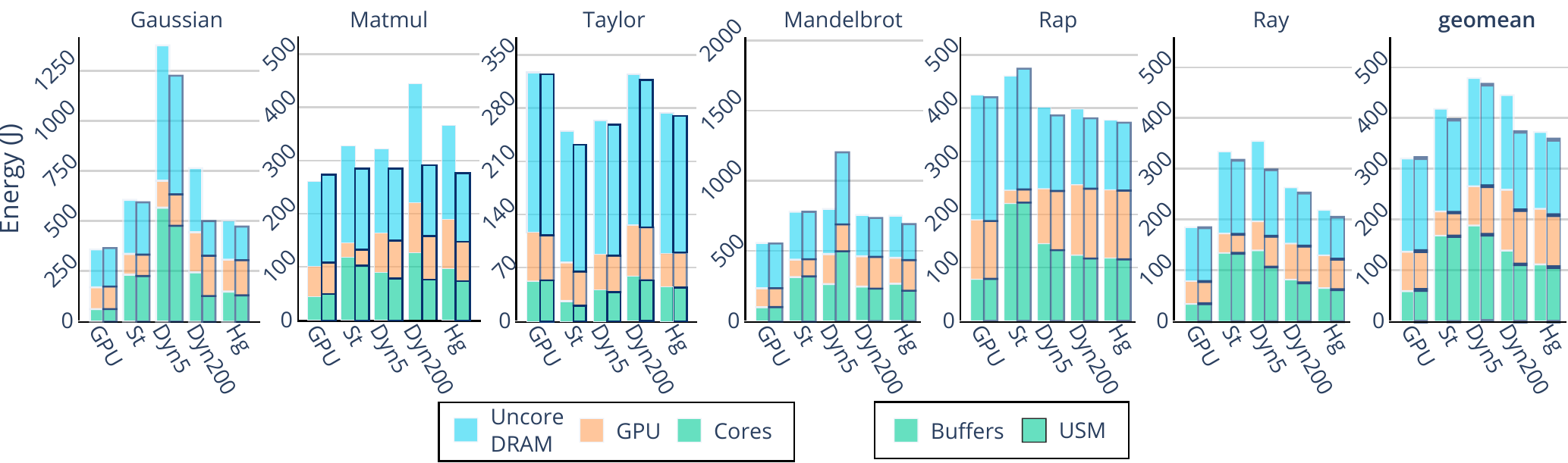}
    \vspace{-2mm}
    \caption{Energy consumption by cores, GPU and the other units of the package with the DRAM consumption.}
    \label{fig:energy-type}
\end{figure*}

\begin{figure*}[!t]
    \centering
    \includegraphics[width=0.99\textwidth, keepaspectratio]{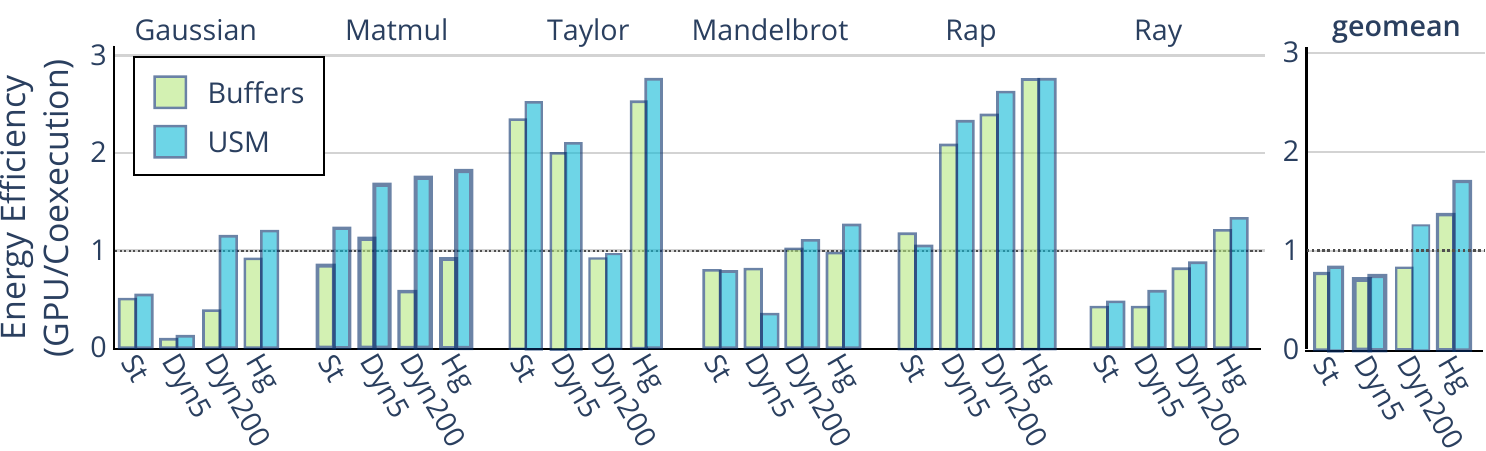}
    \vspace{-2mm}
    \caption{Energy Efficiency compared with GPU (more is better).}
    \label{fig:energy-efficiency}
\end{figure*}







Figure \ref{fig:energy-type} presents the energy consumption, with each bar composed of up to three regions representing the energy used by: the CPU cores, the GPU and the rest of the CPU package together with the DRAM (\emph{uncore + dram}).



Considering the average energy consumption, using only the GPU is the safest option to ensure minimum energy consumption. This is because the energy savings achieved by the reduction in execution time thanks to co-execution, is not enough to counteract the increase in power consumption caused by the use of CPU cores. However, there are also benchmarks such as Taylor and Rap where co-executing does improve power consumption over GPU, and others where co-execution and GPU-only have similar energy consumption, such as MatMul.


Regarding the schedulers, there is a clear correlation between performance and energy consumption. Therefore, the algorithms that offer the best performance in co-execution are also the ones that consume the least energy. On the contrary, the schedulers that cause a lot of imbalance by giving more work to the CPU, spike the energy consumption, due to the higher usage of CPU cores, like Gaussian and Mandelbrot with Dyn5, and RAP with Static.

Another very interesting metric is energy efficiency, which relates performance and energy consumption. In this case it is represented by the ratio of the Energy-Delay Product of the GPU with respect to the co-execution, presented in Figure \ref{fig:energy-efficiency}. Therefore, values higher than 1.0 indicate that the co-execution is more energy efficient than the GPU.

Looking at the geometric mean, it can be concluded that co-execution is 72\% more energy efficient than the GPU execution, using the HGuided scheduler and the USM memory model. Furthermore, this metric is indeed favorable to co-execution in all benchmarks studied, reaching improvements of up to 2.8x in Taylor and RAP.
Thus,
while co-execution consumes more energy in absolute terms on some benchmarks, the reduction in execution time compensates for this extra consumption, resulting in a better performance-energy trade-off.

\begin{figure*}[!t]
  \centering
  \includegraphics[width=0.89\linewidth, keepaspectratio]{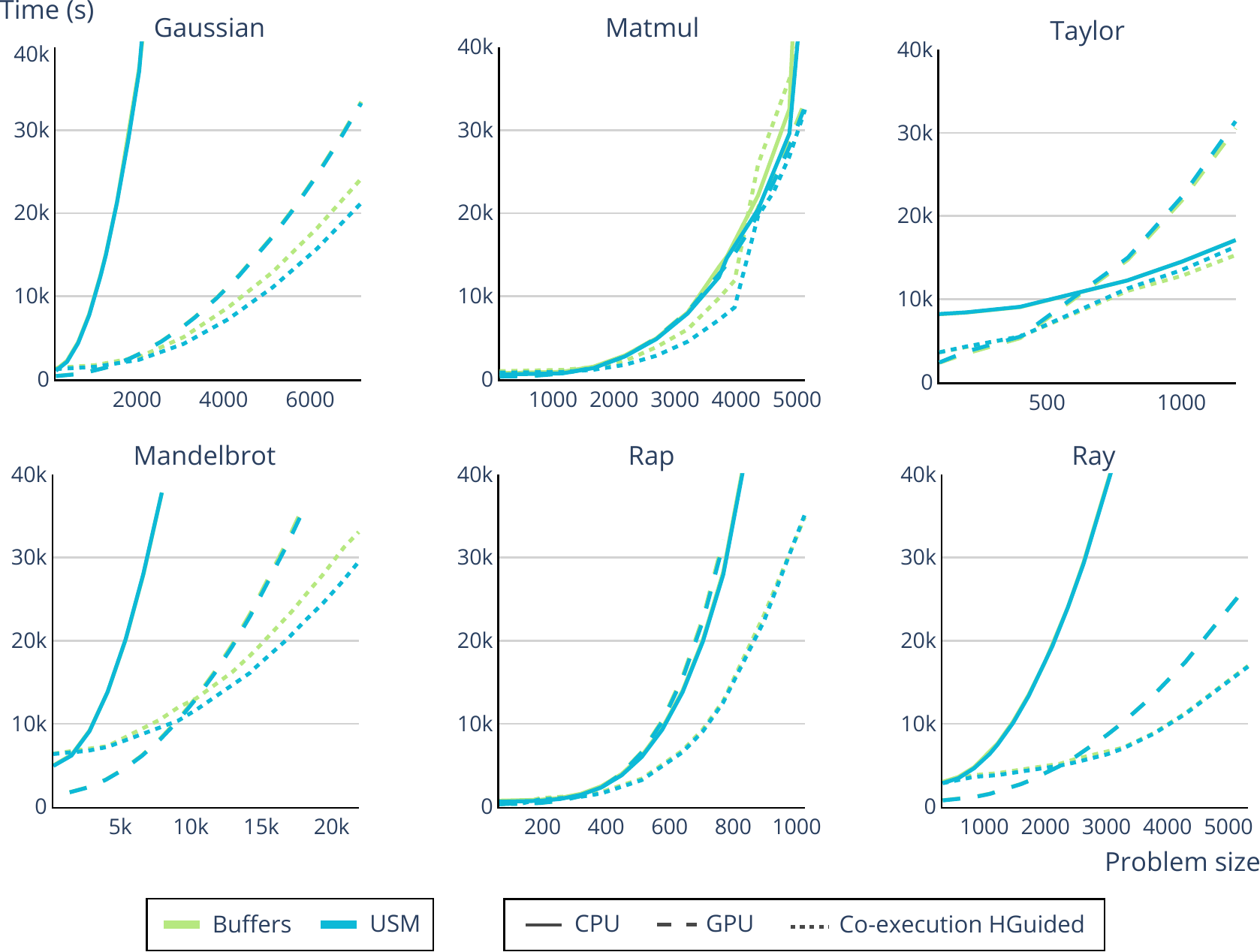}
  \caption{Scalability for CPU, GPU and CPU-GPU coexecution using the Coexecutor runtime with its HGuided scheduling policy.}
  \label{fig:scalability}
\end{figure*}


\subsection{Scalability}

The results presented above refer to problem sizes that need around 10 seconds in the fastest device (GPU). This section presents a scalability analysis of the runtime, varying the size of the problems. To this aim, Figure \ref{fig:scalability} shows the evolution of the execution time of each benchmark with respect to the size of the problem, in different configurations: CPU-only, GPU-only and co-executing. Also, with the two memory models: {\em Buffers} and {\em USM}.

The most important conclusion to be drawn is that, in all the cases studied, there is a turning point from which co-execution improves the performance of the fastest device.
For very small problem sizes, the overhead introduced by the runtime cannot be compensated by the performance increase provided by the co-execution. These points are more noticeable in Gaussian, Mandelbrot and Ray, because the differences in computing capacity between CPUs and GPUs are much more pronounced (13.5x, 4.8x y 4.6x, respectively). Regarding memory models, it is important to note that there are only substantial improvements between {\em USM} and {\em Buffers} in Gaussian, Matmul and Mandelbrot. This is because they are the benchmarks which use more memory, as can be seen in Table \ref{tbl:benchs-properties}. These improvements become greater as the size of the problem increases.

Matmul is a special case, since by increasing the size of the problem, a point is reached where co-execution obtains the same performance as the GPU-only. A detailed analysis of the hardware counters indicates how the LLC memory suffers constant invalidations between CPU and GPU. Temporary locality of the shared memory hierarchy is penalized when co-executing with very large matrices, because the GPU requests memory blocks aggressively.






\section{Conclusions}
\label{sec:Conclusions}


Hardware heterogeneity complicates the development of efficient and portable software, due to the complexity of their architectures and a variety of programming models. In this context, Intel has developed oneAPI, a new and powerful SYCL-based unified programming model with a set of domain-focused libraries, facilitating the development among various hardware architectures.

This paper provides co-execution to oneAPI to squeeze the performance out of heterogeneous systems.
The {\em Coexecutor Runtime} overcomes one of the main challenges in oneAPI, the exploitation of dynamic decisions efficiently. Three load balancing algorithms are implemented on this runtime, showing the behavior in a set of regular and irregular benchmarks.

Furthermore, a validation of performance, balancing efficiency and energy efficiency is carried out, as well as a scalability study. The results indicate that co-execution is worthwhile when using dynamic schedulers, specifically when using HGuided algorithm and unified memory. All achieved due to efficient synchronization, architecture design decisions, computation and communication overlap, and the underlying oneAPI technology and its DPC++ compiler and runtime.



It is important to emphasize that the co-execution has been validated with CPU and integrated GPU, but the proposed runtime is also capable of using other types of architectures that will be incorporated into oneAPI. Therefore, in the future, the co-execution runtime will be extended to evaluate new heterogeneous devices, such as FPGAs and discrete GPUs. In addition, workload schedulers that can take advantage of the benefits offered by this programming model will be designed, thanks to the results presented in this paper.

\section*{Acknowledgment}

This work has been supported by the Spanish Ministry of Education (FPU16/ 03299 grant), the Spanish Science and Technology Commission under contract PID2019-105660RB-C22 and the European HiPEAC Network of Excellence.

\bibliographystyle{plain}
\bibliography{oneapi}
\end{document}